\documentclass[english,english]{article}
\usepackage[T1]{fontenc}
\usepackage[koi8-r]{inputenc}
\usepackage{esint}

\makeatletter
\@ifundefined{date}{}{\date{}}
\newtheorem{theorem}{Theorem}

\newtheorem{lemma}{Lemma}

\newtheorem{remark}{Remark}

\usepackage{babel}

\usepackage{babel}

\makeatother

\usepackage{babel}
\begin{document}

\title{Self-organized circular flow of classical point particles }

\author{V. A. Malyshev %
\thanks{Faculty of Mathematics and Mechanics, Lomonosov Moscow State University,
Moscow, 119991, Russian Federation, E-mail: malyshev2@yahoo.com %
}}
\maketitle
\begin{abstract}
We consider newtonian dynamics of $N$ charged particles on the circle
with nearest neigbour interaction with Coulomb repulsive potential
$r^{-1}$ . Also there is an external accelerating force which is
nonzero only on a small part of the circle. We construct homogeneous
solutions where the velocities of all particles are approximately
equal and their density is approximately uniform. This gives a qualitative
mathematical model for some features of the direct electric current
(DC), in agreement with a suggestion by R. Feynman. 
\end{abstract}

\section{Introduction}

The most developed part of mathematical statistical physics is the
theory of equilibrium (Gibbs) states on discrete lattices. This science
is based on simple axioms followed by many solved problems, including
very difficult. However, in nonequilibrium statistical physics in
continuous space there are problems not even formalized on the mathematical
level.

One of such examples is the direct electric current (DC). On the macro
level it is described by Ohm's law, and, on the micro level, it is
often presented as the classical system of free or weakly interacting
electrons, each accelerated by the constant external force and impeded
by external media. There is an abundance of such models: the first
such model refers to Drude, 1900, one can find it in any textbook
on condensed matter physics, see for example \cite{AshMer}. Recent
papers deeply investigate possible friction mechanisms, see for example
\cite{bruneau,CapMarPul,Spohn}. Here, on the contrary, we ignore
this friction problem but turn to another fundamental problem.

The question arises where the accelerating force comes from, because
among hundred kilometers of power lines the external force acts only
on some meters of the wire. Here what one can read in the ``Feynman
lectures on physics'' (volume 2, sec. 16-2):

``...The force pushes the electrons along the wire. But why does
this move the galvanometer, which is so far from the force ? Because
when the electrons which feel the magnetic force try to move, they
push - by electric repulsion - the electrons a little farther down
the wire; they, in turn, repel the electrons a little farther on,
and so on for a long distance. An amazing thing. It was so amazing
to Gauss and Weber - who first built a galvanometer - that they tried
to see how far the forces in the wire would go. They strung the wire
all the way across the city...''.

This was written by the famous physicist. However, after that, this
``amazing thing'' was vastly ignored in the literature. Many more
questions arise. For example, why DC moves slowly but its stationary
regime is being established almost immediately. Here we are occupied
with the first one (some results concerning the second problem see
in \cite{Mal_analytic_4}). Namely, we want to demonstrate rigorously
that even on the classical (non-quantum) level there is a mere possibility
that the stationary and space homogeneous flow of charged particles
may exist as a result of self-organization of strongly interacting
(via Coulomb repulsion) system of electrons.

We use only classical nonrelativistic physics - Newtonian dynamics
and Coulomb's law, but also the simplest friction mechanism, ignoring
where this friction mechanism comes from. Rigorous models of strongly
interacting electron systems, interacting also with the ionic lattice,
do not exist now.

\paragraph{The Model}

We consider $N$ point particles $i=1,2,...,N$ initially at the points
\[
0\leq x_{1}(0)<...<x_{N}(0)<L
\]
of the interval $[0,L]\in R$. We assume periodic boundary conditions
that is we consider the circle $S_{L}=[0,L)$ of length $L$. The
trajectories $x_{i}(t)$ are defined by the following system of $N$
equations

\begin{equation}
M\frac{d^{2}x_{i}}{dt^{2}}=-\frac{\partial U}{\partial x_{i}}+gF(x_{i})-a(\frac{dx_{i}}{dt})\label{main_eq}
\end{equation}
The interaction $U$ between the particles is

\[
U(\{x_{i}\})=\sum_{i=1}^{N}V(x_{i+1}-x_{i})
\]
where of course $x_{N+1}=x_{1}$ and 
\[
V(x)=V(-x)=\frac{\alpha}{r}>0,r=|x|
\]
The case $\alpha>0$ corresponds to the Coulomb repulsive potential,
that we consider here. Then the repulsive force is 
\[
f(r)=-\frac{dV(r)}{dr}=\alpha r^{-2}
\]
It follows that the particles, during the movement, cannot change
their order. $gF(x)$ - an external accelerating force (assumed sufficiently
smooth) with scaling parameter $g>0$. The friction function $a(v)$
is specified below. It defines the loss of kinetic energy via the
interaction with external media.

It is well-known that the solution of the system (\ref{main_eq}),
for any initial conditions, exists and is unique on all time interval
$[0.\infty)$, under sufficiently general assumptions on the functions
$F$ and $a$. However, to get more detailed information about trajectories,
one needs sufficient efforts.

\section{Effective force}

\paragraph{Parameters and constants}

Throughout the paper we are dealing with macro and micro parameters
and (absolute) constants. Absolute constants do not depend on the
parameters of the model.

All our constructions are for $N$ sufficiently large but finite -
one cannot directly perform the limit $N\to\infty$ because there
will be different micro-scales, influencing on the macro-parameters.
Macro-parameters $L$ and $F(x)$ are fixed (do not depend on $N$),
for example we put 
\[
C(F,m)=\max_{x}|F^{(m)}(x)|,m=0,1,2,
\]
and put for convenience 
\[
C(F,0)=1
\]
Roughly speaking, our first approximation dynamics is 
\[
x_{k}(t)=x_{k}(0)+Vt,\frac{dx_{k}(t)}{dt}=V
\]
where $V$ is a macro-parameter (the approximate velocity of particles).

Micro-parameters 
\[
M=M^{(N)},g=g^{(N)},\alpha=\alpha^{(N)}
\]
depend on $N$ (they satisfy some conditions defined below), but we
will omit index $(N)$. For example, in some physical situation (in
SI units) approximately 
\[
N^{-1}=10^{-10},M=10^{-30},\alpha=10^{-28}
\]
These numbers were some guide for us, but we could not fit them completely
- $g$ had to be assumed smaller than necessary.

\paragraph{Static configurations}

For any particle configuration $(x_{i}=x_{i}^{(N)},v_{i}=v_{i}^{(N)})$
\[
0\leq x_{1}=x_{1}^{(N)}<...<x_{N}=x_{N}^{(N)}<L
\]
the \textbf{effective} force, acting on the particle $i$, is 
\[
w_{i}-a(v_{i})
\]
where

\[
w_{i}=w(x_{i-1},x_{i},x_{i+1})=f(x_{i}-x_{i-1})-f(x_{i+1}-x_{i})+gF(x_{i})
\]
The following crucial result depends only on the parameter $C_{\alpha,g}=\alpha^{-1}g$.

\begin{lemma}\label{lemma_w}

Assume $C_{\alpha,g}$ fixed or bounded as $N\to\infty$, then for
any sufficiently large $N$ there exists configuration $x_{1}<...<x_{N}$
(assuming zero velocities) such that the effective force is the same
for all $i=1,...,N$, that is there exists $w$ such that 
\begin{equation}
f(x_{i}-x_{i-1})-f(x_{i+1}-x_{i})+gF(x_{i})=w_{i}=w,i=1,...,N\label{Equat_constantForce}
\end{equation}
Moreover, for this configuration the following properties hold: 
\begin{enumerate}
\item uniformly in $i=1,...,N$ as $N\to\infty$ 
\[
\Delta_{i}=x_{i+1}-x_{i}\sim\frac{L}{N}
\]

\item for any $i$ denote $\Delta_{i}=\Delta(1+\delta_{i}),\Delta_{1}=\Delta$.
Then for all $i$ 
\[
|\delta_{i}|\leq4\alpha^{-1}gL_{0}\Delta
\]
where $L_{0}$ is the length of the support of $F(x)$. 
\end{enumerate}
\end{lemma}

Proof. Let us call $\psi(x)=gF(x)-w$ the virtual force. Then the
required configuration can be interpreted as a fixed point of $N$
particle system for the external virtual force. The virtual force
is potential iff 
\begin{equation}
\int_{S_{L}}\psi(x)dx=g\int_{S_{L}}F(x)dx-Lw=0\label{cycle-1}
\end{equation}
Then the (virtual) potential of the virtual force is 
\[
W(x)=-\int_{0}^{x}\psi(x)dx
\]
and a required fixed point exists as a global minimum (in $R^{N}$)
of the potential 
\[
U(x_{1},...,x_{N})+\sum_{i=1}^{N}W(x_{i})
\]
If such minimum is not unique, we take anyone.

The effective force $w>0$ can be found from condition (\ref{cycle-1}).
At the same time, summing up the equations (\ref{Equat_constantForce}),
we get 
\[
g\sum_{i=1}^{N}F(x_{i})-Nw=0
\]
Thus, the constant effective force equals 
\begin{equation}
w=g\frac{1}{L}\int_{S_{L}}F(x)dx=g\frac{1}{N}\sum_{i=1}^{N}F(x_{i})\label{cycle-2}
\end{equation}

Assertion 1 of the Lemma was proved in Theorem 1 of \cite{Mal_fixed_3},
see also \cite{Mal_fixed_2}. Now let us prove the assertion 2 of
the lemma. Summing up the equations (\ref{Equat_constantForce}) for
$i=2,...,k$, we get 
\[
f(\Delta(1+\delta_{k}))-f(\Delta)=\sum_{i=2}^{k}(gF(x_{i})-w)
\]
or 
\[
(1+\delta_{k})^{-2}-1=\alpha^{-1}\Delta^{2}\sum_{i=2}^{k}(gF(x_{i})-w)
\]
Then 
\begin{equation}
\delta_{k}=[1+Q_{k}]^{-\frac{1}{2}}-1=\sum_{m=1}^{\infty}a_{m}Q_{k}^{m},Q_{k}=\alpha^{-1}\Delta^{2}\sum_{i=2}^{k}(gF(x_{i})-w)\label{delta_k}
\end{equation}

\begin{equation}
(-1)^{m}a_{m}=\frac{1.3...(2m-1)}{2^{m}m!}=\frac{(2m)!}{(2^{m}m!)^{2}}\sim\frac{1}{\sqrt{\pi m}},|a_{m}|\leq1\label{a_m}
\end{equation}
and 
\begin{equation}
|w|\leq g\frac{L_{0}}{L},\,\,\,|\sum_{i=2}^{k}(gF(x_{i})-w)|\leq(1+\frac{L_{0}}{L})g(k-1)\label{w_estimate}
\end{equation}
It follows that for sufficiently large $N$ 
\[
|\delta_{k}|\leq\sum_{m=1}(\alpha^{-1}(1+\frac{L_{0}}{L})gL_{0}\Delta)^{m}=\frac{\alpha^{-1}(1+\frac{L_{0}}{L})gL_{0}\Delta}{1-\alpha^{-1}(1+\frac{L_{0}}{L})gL_{0}\Delta}\leq2(1+\frac{L_{0}}{L})\alpha^{-1}gL_{0}\Delta
\]
if $\alpha^{-1}(1+\frac{L_{0}}{L})gL_{0}\Delta\leq\frac{1}{2}$. The
Lemma is proved.

\begin{remark} The assertion 1 of Lemma \ref{lemma_w} says that
the asymptotics of $\Delta_{k}$ does not depend on $F$. We say in
this case that $F$ is not seen on the microscale $N^{-1}$ but only
on the sub-microscale, see \cite{Mal_fixed_2} and \cite{Mal_fixed_3}.
It follows that the density is macro-homogeneous. However, as it is
clear from (\ref{delta_k}) and (\ref{a_m}), on the interval where
$F(x)=0$, the distances between particles slightly increase in the
clock-wise direction.

\end{remark}

\section{Main result - macro-homogeneous dynamics}

For any sufficiently large $N$ we shall prove existence of the dynamics
for $t\in[0,T),T=T(N)\approx N$, which we call \textbf{macro-homogeneous}
on $[0,T)$. That is we shall prove that the following two properties
hold as $N\to\infty$: 
\begin{enumerate}
\item (asymptotically homogeneous velocities) There exists constant (one
more macro-parameter) $V>0$ such that uniformly in $i=1,2...,N$
and $t\in[0.T)$ the velocities 
\[
v_{i}(t)=v_{i}^{(N)}(t)\to V
\]

\item (asymptotically homogeneous density) uniformly in $t\in[0.T)$, for
any interval $I\subset S_{L}$ the number $N(I,t)$ of particles in
$I$ 
\[
N(I,t)\sim\frac{|I|}{L}N
\]

\end{enumerate}

\paragraph{Assumptions}

We choose the simplest friction mechanism defined by the function
which is linear in the (micro) vicinity of the macroparameter $V$
\[
a(v)=A_{0}+Av,v\in(V-\Delta,V+\Delta)
\]
where $A_{0}$ and $A>0$ are microparameters, coordinated with macroparameter
$V$ and microparameter $w$ so that 
\begin{equation}
a(V)=A_{0}+AV=w>0\label{a_w}
\end{equation}
Concerning the parameters, roughly speaking, there are two assumptions:
$g$ is small enough, in particular $g\ll A$ and 
\[
N^{-1}A^{2}\alpha^{-1}\ll M\ll\min(\alpha N^{-1},A)
\]
More exactly, it can be formulated as follows. For any sufficiently
large $N$ and some sufficiently small absolute constant $\rho>0$
\begin{equation}
N^{-1}A^{2}\alpha^{-1}\rho^{-1}<M<\rho\min(\alpha N^{-1}\frac{1}{\ln N},A)\label{cond_main_1}
\end{equation}
\begin{equation}
C_{A,g}=A^{-1}g\leq\rho N^{-3}\label{cond_main_2}
\end{equation}

In particular, we will use below the first inequality in (\ref{cond_main_1})
more concretely 
\begin{equation}
\frac{M\alpha N}{A^{2}}>(16(2\pi)^{2})^{-1}\label{cond_roots}
\end{equation}
The following example shows that these inequalities provide non-empty
and natural domain of parameters 
\[
g=N^{-\gamma_{1}},\alpha=N^{-\gamma_{2}}.M=N^{-\gamma_{3}}
\]
with constants $\gamma_{1}>2,\gamma_{2}>0,\gamma_{3}>0$ and such
that 
\[
\gamma_{1}>\gamma_{2},1+2\gamma_{1}-\gamma_{2}>\gamma_{3}>\max(1+\gamma_{2},\gamma_{1})
\]
Put also 
\begin{equation}
C_{w,g}=w^{-1}g=\frac{L}{C(F,int)},C(F,int)=\int F(x)dx\label{condition_w_g}
\end{equation}

\paragraph{Initial conditions}

We fix the initial configuration for the required dynamics as $v_{i}=V$
and $x_{i}$ are chosen as in Lemma \ref{lemma_w}, that is 
\[
x_{k+1}(0)-x_{k}(0)=\Delta_{k},w=\alpha\Delta_{k-1}^{-2}-\alpha\Delta_{k}^{-2}+gF(x_{k}(0))
\]
and it is convenient to choose the coordinate system by $x_{1}(0)=0$.
Thus the initial effective force is zero.

We could choose the initial velocities as 
\[
v_{k}(0)=V+u_{k}(0)
\]
where $u_{k}(0)$ could be assumed sufficiently small, but for convenience
we always assume $u_{k}(0)=0$.

\paragraph{Main result}

\begin{theorem}\label{theorem_main}

Assume (\ref{a_w}) and (\ref{cond_main_1})-(\ref{cond_main_2}).
Then for the chosen initial conditions (explicitely defined in the
next section) there exists $\beta>0$, not depending on $N$, such
that the dynamics is macro-homogeneous on the time interval $[0.T),T=T_{N}=\beta N$.

\end{theorem}

\paragraph{Plan of the proof}

Choosing initial configuration $x_{k}(0)$ is a delicate matter. We
did it in the previous section so that the effective forces acting
on each particle are the same at time $0$. This dynamics does not
satisfy the main equations and we derive (in section 4) the equations
for the deviations $y_{k}(t)=x_{k}(t)-x_{k}(0)-Vt$. It is very important
that our choice of the initial conditions allows to exclude the constant
component of the effective force in the equations for $y_{k}(t)$.

We fix the basic linear part of these equations for $y_{k}(t)$ and
solve them (section 5). It is rather straightforward but demands delicate
estimates. The rest linear and nonlinear parts of the equations are
considered as the perturbation and demand some iteration procedure.
The section 6 is devoted to convergence of this procedure and to stability
estimates. To prove this we introduce special Banach space where the
convergence holds, that also demands some nontrivial estimates. There
are many similarities in the estimates - we tried not to repeat them.
Finally, in section 7, we give some remarks and perspective.

In the rigorous proof we tried to be very accurate with micro-parameters,
that is with the parameters depending on $N$. Absolute constants
(we will meet finite number of them) are denoted $c,c_{1},c_{2},...$
and could be easily explicitely written but we did not do this because
of no interest. Also sometimes we denote $C(F)$ a generic macro-constant
depending only on $F$. Starting from section 5 we take $L=1$.

\section{Equations for the deviations}

As the force $F(x)$ is not translation invariant, the dynamics $x_{k}(t)=x_{k}(0)+Vt$.
cannot satisfy equations (\ref{main_eq}). We introduce the deviations
$y_{k}(t)$ and their velocities $u_{k}(t)$ by 
\[
x_{k}(t)=x_{k}(0)+Vt+y_{k}(t),y_{k}(0)=0,
\]

\[
v_{k}(t)=\frac{dx_{k}}{dt},u_{k}(t)=\frac{dy_{k}}{dt}=v_{k}(t)-V,y_{k}(t)=\int_{0}^{t}u_{k}(t)dt,
\]
and rewrite the main equations (\ref{main_eq}) as equations for the
deviations

\[
M\frac{d^{2}y_{k}}{dt^{2}}=f(\Delta_{k-1}(0)+y_{k}(t)-y_{k-1}(t))-f(\Delta_{k}(0)+y_{k+1}(t)-y_{k}(t))+
\]

\begin{equation}
+gF(x_{k}(0)+Vt+y_{k}(t))-a(V+u_{k}(t))\label{main_deviations}
\end{equation}

For any function $h(k)$ on the finite cyclic group $\{k:k=1,...N_{L}\}$,
with zero element $N_{L}=0$, introduce the shift operator 
\[
(Sh)(k)=h(k+1)
\]
and finite difference operators (discrete derivatives)

\[
\nabla_{k}h(k)=\nabla h(k)=\nabla^{+}h(k)=h(k+1)-h(k),(\nabla^{-}h)(k)=h(k)-h(k-1),
\]
in particular 
\[
\nabla y_{k}(t)=y_{k+1}(t)-y_{k}(t)
\]
Sometimes we will use Leibnitz formula for discrete derivatives (concerning
calculus of finite differences see \cite{Mal_fine} and references
to classical papers therein)

\begin{equation}
\nabla(f_{1}...f_{n})=(\nabla f_{1})S(f_{2}...f_{n})+f_{1}\nabla(f_{2}...f_{n})=...=\sum_{m=1}^{n}f_{1}...f_{m-1}(\nabla f_{m})S(f_{m+1}...f_{n})\label{Leibnitz}
\end{equation}
Note that it differs from the standard Leibnitz formula for differentiation
only by shift operators. As we will use it only for estimates from
above, which are always uniform in $k$, the shift operators will
not play role.

Then 
\[
f(\Delta_{k-1}(0)+y_{k}(t)-y_{k-1}(t))-f(\Delta_{k}(0)+y_{k+1}(t)-y_{k}(t))=\alpha(-\nabla^{-})(\Delta_{k}(0)+\nabla^{+}y_{k})^{-2}=
\]

\begin{equation}
=\alpha(-\nabla^{-})(\Delta_{k}^{-2}-2\frac{\nabla^{+}y_{k}(t)}{\Delta_{k}^{3}})+L_{3}\label{L_3}
\end{equation}

\[
L_{3}=\alpha(-\nabla^{-})[(\Delta_{k}(0)+\nabla^{+}y_{k})^{-2}-\Delta_{k}^{-2}+2\frac{\nabla^{+}y_{k}(t)}{\Delta_{k}^{3}}]
\]
Formally 
\begin{equation}
L_{3}=\alpha(-\nabla^{-})\Delta_{k}^{-2}\sum_{m=2}^{\infty}d_{m}(\frac{\nabla^{+}y_{k}(t)}{\Delta_{k}})^{m},d_{m}=(-1)^{m}(m+1)\label{L_3_series}
\end{equation}
Applying $-\alpha\nabla^{-}$ to the first and second terms in the
right-hand side of (\ref{L_3}), and putting 
\[
\delta_{k}^{1}=(1+\delta_{k})^{-3}-1
\]
we have 
\[
-\alpha\nabla^{-}\Delta_{k}^{-2}=-\alpha(\Delta_{k}^{-2}-\Delta_{k-1}^{-2}),
\]

\[
\alpha(-\nabla^{-})[-2\Delta_{k}^{-3}\nabla^{+}y_{k}(t)]=2\alpha[\Delta_{k}^{-3}\nabla^{+}y_{k}(t)-\Delta_{k-1}^{-3}\nabla^{+}y_{k-1}(t))]=
\]

\[
=2\alpha\Delta^{-3}[(1+\delta_{k})^{-3}(y_{k+1}-y_{k})-(1+\delta_{k-1})^{-3}(y_{k}-y_{k-1})]=2\alpha\Delta^{-3}[y_{k+1}(t)-2y_{k}(t)+y_{k-1}(t)]+L_{2}
\]
where 
\[
L_{2}=2\alpha\Delta^{-3}\nabla^{-}[\delta_{k}^{1}\nabla y_{k}]
\]
Then the main equations become 
\[
M\frac{d^{2}y_{k}}{dt^{2}}+A\frac{dy_{k}(t)}{dt}+[\alpha\Delta_{k}^{-2}-\alpha\Delta_{k-1}^{-2}-gF(x_{k}(0))+w]-2\alpha\Delta^{-3}[y_{k+1}-2y_{k}+y_{k-1})]=
\]

\[
=L_{2}+L_{3}+gF(x_{k}(0)+Vt+y_{k}(t))-gF(x_{k}(0))-A\frac{dy_{k}}{dt})
\]
As the first square bracket (that is the ``effective acceleration
force'') is zero, we get the final form of the (differential-difference)
equations

\begin{equation}
M\frac{d^{2}y_{k}}{dt^{2}}+A\frac{dy_{k}(t)}{dt}-2\alpha\Delta^{-3}[y_{k+1}-2y_{k}+y_{k-1})]=L_{1}+L_{2}+L_{3}+g\phi_{k}(t)\label{main_nonlinear}
\end{equation}
where 
\[
\phi_{k}(t)=F(x_{k}(0)+Vt)-F(x_{k}(0))
\]

\[
L_{1}=g[F(x_{k}(0)+Vt+y_{k}(t))-F(x_{k}(0)+Vt)]
\]
First of all, we shall study in detail a cut-off system 
\begin{equation}
M\frac{d^{2}y_{k}(t)}{dt^{2}}+A\frac{dy_{k}(t)}{dt}-2\alpha\Delta^{-3}[y_{k+1}-2y_{k}+y_{k-1})]=g\phi_{k}(t)\label{zeroLinear}
\end{equation}
which we call the basic linear approximation and consider $L_{i},i=1,2,3$
as perturbation terms. We take them into account in section 6.

\section{Linear stability}

From now on we put $L=1,N=N_{1}$. Denote $y_{k,0}(t)$ the solution
of equations (\ref{zeroLinear}).

\subsection{Fourier transform}

Denote 
\begin{equation}
\Phi(h)=\Phi(h)(n)=\frac{1}{N}\sum_{k=1}^{N}h_{k}\exp(2\pi in\frac{k}{N})\label{fourier}
\end{equation}
the Fourier transform 
\[
\Phi:l_{2}(\{\frac{k}{N}:k=1,...,N\})\to l_{2}(\{n=0,1,...,N-1\})
\]
of the function $h=h_{k}$ on $\{\frac{k}{N}:k=1,...,N\}$. The inverse
Fourier transform is 
\begin{equation}
h_{k}=\sum_{n}\Phi(h)(n)\exp(-2\pi in\frac{k}{N})\label{inverseFourier}
\end{equation}
Put 
\[
\eta(n,t)=\Phi(y_{k,0}(t)),\tilde{\phi}(n,t)=\Phi(\phi_{k}(t))
\]
Multiplying (\ref{zeroLinear}) on $\frac{1}{N}\exp(2\pi in\frac{k}{N})$,
summing in $k$ and dividing by $M$, we get the decoupled equations
for $n=0,1,...,N-1$

\begin{equation}
\frac{d^{2}\eta(n,t)}{dt^{2}}+\frac{A}{M}\frac{d\eta(n,t)}{dt}+\frac{4\alpha}{M}\Delta^{-3}(1-\cos2\pi n\frac{1}{N})\eta(n,t)=M^{-1}g\tilde{\phi}(n,t)\label{fourier_Linearequation}
\end{equation}

The characteristic equation 
\[
Q(z)=z^{2}+\frac{A}{M}z+\frac{4\alpha}{M}\Delta^{-3}(1-\cos2\pi n\frac{1}{N})=0
\]
has the roots 
\begin{equation}
z_{1,2}(n)=-\frac{A}{2M}(1\pm\sqrt{1-\frac{16M\alpha}{A^{2}\Delta^{3}}(1-\cos2\pi n\frac{1}{N})})\label{z_1_2}
\end{equation}
If 
\[
1-\frac{16M\alpha}{A^{2}\Delta^{3}}(1-\cos2\pi n\frac{1}{N})\neq0
\]
then the roots are different and the general solution is 
\[
\eta(n,t)=\eta_{1}(n,t)+\eta_{2}(n,t)
\]
where for $l=1,2$ 
\[
\eta_{l}(n,t)=C_{l}(n)e^{z_{l}t}+\frac{e^{z_{l}t}}{Q'(z_{l})}\int_{0}^{t}e^{-z_{l}t_{1}}M^{-1}g\tilde{\phi}(n,t_{1})dt_{1}=
\]

\begin{equation}
=C_{l}(n)e^{z_{l}t}+\frac{M^{-1}}{2z_{l}+\frac{A}{M}}\int_{0}^{t}e^{z_{l}(t-t_{1})}g\tilde{\phi}(n,t_{1})dt_{1}=C_{l}(n)e^{z_{l}t}+r_{l}(n)g\int_{0}^{t}e^{z_{l}(t-t_{1})}\tilde{\phi}(n,t_{1})dt_{1}\label{eta_0}
\end{equation}
where 
\begin{equation}
r_{l}(n)=(-1)^{l}\frac{1}{\frac{A}{2}\sqrt{1-\frac{16M\alpha}{A^{2}\Delta^{3}}(1-\cos2\pi n\frac{1}{N})})}\label{r_l}
\end{equation}
As $y_{k}(0)=0$ we have 
\[
C_{2}=-C_{1},
\]
Introduce the Fourier transform of $u_{k}(0)$ 
\[
\eta'(n,0)=\frac{\partial\eta(n,0)}{\partial t}=\sum_{l=1,2}C_{l}(n)z_{l}(n)=C_{1}(n)(z_{1}-z_{2})
\]
It follows 
\[
|C_{1}|=|C_{2}|=|\frac{\eta'(n)}{z_{1}-z_{2}}|\leq\frac{M}{A}|\eta'(n,0)|
\]
From now on we assume for simplicity that $u_{k}(0)=0$ for all $k$,
then 
\[
C_{1}=C_{2}=0
\]

\subsection{Main ``linear'' lemmas}

If 
\[
\frac{16M\alpha}{A^{2}\Delta}(2\pi)^{2}>2
\]
(garantied by the left inequality of (\ref{cond_main_1})) then for
any $n\neq0$ the roots $z_{1,2}(n)$ are complex conjugate.

In the linear case we need only boundedness of $C_{A,g}$ and $C_{\alpha,g}$
and condition (\ref{cond_main_1}). Under these conditions we will
prove

\begin{lemma}\label{lemma_eta-1}

\begin{equation}
|\eta(0,t)|\leq\Delta^{2}(t+1)C_{A,g}(C(F)+4C_{\alpha,g})\label{complex_n_0}
\end{equation}
and for any $n\neq0$ 
\begin{equation}
|\eta(n,t)|\leq\frac{c}{n}C_{A,g}\sqrt{\frac{M\Delta}{\alpha}}\label{complex_n_neq_0}
\end{equation}
\end{lemma}

Proof. We have for $n\neq0$ 
\[
\Re z_{1}(n)=-\frac{A}{2M},
\]

\[
\eta_{l}(n,t)=r_{l}(n)\int_{0}^{t}e^{z_{l}(t-t_{1})}g\tilde{\phi}(n,t_{1})dt_{1}
\]
As, by the left inequality in (\ref{cond_main_1}), $\frac{M\alpha}{A^{2}\Delta}$
is sufficiently large, and $1-\cos2\pi n\frac{1}{N}>c_{1}(\frac{n}{N})^{2}$
for some $c_{1}>0$ and any $n\neq0$, then for $l=1,2$ from (\ref{r_l})
we have 
\[
g|r_{l}(n)|=2C_{A,g}|\frac{1}{\sqrt{1-\frac{16M\alpha}{A^{2}\Delta^{3}}(1-\cos2\pi n\frac{1}{N})})}|\leq\frac{2C_{A,g}}{\sqrt{\frac{16M\alpha}{A^{2}\Delta^{3}}(\frac{n}{N})^{2}c_{1}}}
\]
Then 
\[
|\eta_{1}(n,t)|\leq\frac{2C_{A,g}}{\sqrt{\frac{16M\alpha}{A^{2}\Delta^{3}}(\frac{n}{N})^{2}c_{1}}}|\int_{0}^{t}e^{z_{1}(t-t_{1})}dt_{1}|=\frac{2C_{A,g}}{\sqrt{\frac{16M\alpha}{A^{2}\Delta^{3}}(\frac{n}{N})^{2}c_{1}}}|\Re z_{1}|^{-1}(1-e^{\Re z_{1}t})\leq
\]

\begin{equation}
\leq c\frac{1}{n}C_{A,g}\sqrt{\frac{M\Delta}{\alpha}}\label{eta_1_n_linear}
\end{equation}
For $n=0$ we have by (\ref{z_1_2}) and (\ref{r_l}) 
\begin{equation}
z_{1}(0)=-\frac{A}{M},z_{2}(0)=0,|r_{1,2}(0)|=\frac{2}{A}\label{linear_zero_1}
\end{equation}
and we need the following

\begin{lemma}\label{lemma_fi_0}

For $n=0$ we have

\begin{equation}
\sup_{t}|\tilde{\phi}(0,t)|\leq\Delta^{2}C(F,2)+8C(F,1)C_{\alpha,g}\Delta\label{lemma_fi_1}
\end{equation}

\begin{equation}
|\int_{0}^{t}\tilde{\phi}(0,t_{1})dt_{1}|\leq\Delta^{2}(2(t+1)(C_{\alpha,g}+C(F,2))+2C(F,1))\label{lemma_fi_2}
\end{equation}

\end{lemma}

From (\ref{linear_zero_1}) and this lemma we get 
\[
|\eta_{1}(0,t)|=g|r_{1}(0)||\Re z_{1}(0)|^{-1}\sup|\tilde{\phi}(0,t)|\leq2C_{A,g}\frac{M}{A}(\Delta^{2}C(F,2)+8C(F,1)C_{\alpha,g}\Delta)\leq
\]

\[
\leq2C_{A,g}(\frac{M}{A}\Delta^{2}C(F,2)+8C(F,1)\Delta^{2})
\]
as $\frac{M}{A}C_{\alpha,g}<\Delta$ by (\ref{cond_main_1}). Also
\[
|\eta_{2}(0,t)|=g|r_{l}(0)\int_{0}^{t}\tilde{\phi}(0,t_{1})dt_{1}\}\leq2C_{A,g}\Delta^{2}(2(t+1)(C_{\alpha,g}+C(F,2))+2C(F,1))
\]

\[
|\eta(0,t)|\leq|\eta_{1}(0,t)|+|\eta_{2}(0,t)|\leq\Delta^{2}(t+1)C_{A,g}(C(F)+4C_{\alpha,g})
\]
where 
\[
C(F)=6C(F,2)+24C(F,1)
\]

Now we can prove the following

\begin{lemma}\label{lemma_y_k_0}

Uniformly in $k$ 
\[
|y_{k,0}(t)|\leq cC_{A,g}\sqrt{\frac{M\Delta}{\alpha}}\ln N+\Delta^{2}(t+1)C_{A,g}(C(F)+4C_{\alpha,g})
\]

\[
|\nabla y_{k,0}(t)|\leq cC_{A,g}\sqrt{\frac{M\Delta}{\alpha}}\ln N,
\]

\[
|\nabla^{2}y_{k+1,0}(t)|\leq cC_{A,g}\sqrt{\frac{M\Delta}{\alpha}}\ln N
\]

\end{lemma}

This is the direct calculation via Fourier transform. We use inverse
Fourier transform (\ref{inverseFourier}), thus we have only to sum
up the terms of Lemma \ref{lemma_eta-1}, this gives 
\[
|y_{k,0}(t)|\leq\sum_{n}\sum_{l=1,2}|\eta_{l}(n)|
\]
We see that the main term corresponds to the zero mode. Noting that
\begin{equation}
\Phi(\nabla y_{k,0}(t))=\eta(n,t)(e^{-2\pi i\frac{n}{N}}-1),\Phi(\nabla^{2}y_{k,0}(t))=\eta(n,t)(e^{-2\pi i\frac{n}{N}}-1)^{2}\label{derivatives_Fourier}
\end{equation}
and 
\begin{equation}
|\nabla y_{k,0}(t)|\leq\sum_{n\neq0}\sum_{l=1,2}|\eta_{l}(n,t)(e^{-2\pi i\frac{n}{N}}-1)|\label{_y_eta_1}
\end{equation}

\begin{equation}
|\nabla^{2}y_{k,0}(t)|\leq\sum_{n\neq0}\sum_{l=1,2}|\eta_{l}(n,t)(e^{-2\pi i\frac{n}{N}}-1)^{2}|\label{y_eta_2}
\end{equation}
We see that for the first and second differences the zero mode vanioshes,
that gives better estimate. The Lemma is proved.

\subsection{Proof of Lemma \ref{lemma_fi_0}}

To prove (\ref{lemma_fi_1}) note that for any $t$ there is $m(1,t)$
such that 
\[
x_{m(1,t)}(0)\leq x_{1}(0)+Vt<x_{m(1,t)+1}(0)
\]
and denote 
\[
\gamma=x_{1}(0)+Vt-x_{m(1,t)}(0)=Vt-x_{m(1,t)}(0)
\]
Put $m(k,t)=m(1,t)+(k-1)$. Then 
\begin{equation}
x_{k}(0)+Vt-x_{m(k,t)}(0)=\gamma+\eta_{k}\label{gamma_plus_feta}
\end{equation}
and uniformly in $k$ 
\[
|\eta_{k}|\leq8C_{\alpha,g}\Delta
\]
This follows from (\ref{delta_k}) and the evident formulas 
\[
x_{k}(0)=x_{1}(0)+\sum_{i=1}^{k-1}\Delta_{i}=\sum_{i=1}^{k-1}\Delta(1+\delta_{i})=(k-1)\Delta+\Delta\sum_{i=1}^{k-1}\delta_{i}
\]

\[
x_{m(k,t)}(0)=x_{m(1,t)}(0)+(k-1)\Delta+\Delta\sum_{i=m(1,t)}^{m(k,t)-1}\delta_{i}
\]

\[
x_{k}(0)+Vt-x_{m(k,t)}(0)=\gamma+\Delta(\sum_{i=m(1,t)}^{m(k,t)-1}\delta_{i}-\sum_{i=1}^{k-1}\delta_{i})
\]
Then we can write 
\[
\tilde{\phi}(0,t)=\frac{1}{N}\sum_{k}(F(x_{k}(0)+Vt)-F(x_{k}(0))=\frac{1}{N}\sum_{k}(F(x_{k}(0)+Vt)-F(x_{m(k,t)}(0))=
\]

\[
=\frac{1}{N}\sum_{k}F^{(1)}(x_{m(k,t)}(0)+\theta_{k})(\gamma+\eta_{k})
\]
where $|\theta_{k}|\leq\gamma+|\eta_{k}|$ and we used (\ref{gamma_plus_feta}).
But 
\[
\frac{1}{N}\sum_{k}F^{(1)}(x_{m(k,t)}(0)+\theta_{k})\to_{N\to\infty}\int_{S_{1}}F^{(1)}(x)dx=0
\]
and moreover 
\[
|\frac{1}{N}\sum_{k}F^{(1)}(x_{m(k,t)}(0)+\theta_{k})|\gamma=|\frac{1}{N}\sum_{k}F^{(1)}(x_{m(k,t)}(0)+\theta_{k})-\int_{S_{1}}F^{(1)}(x)dx|\gamma\leq
\]

\[
\leq C(F,2)\Delta\gamma\leq C(F,2)\Delta^{2}
\]

\[
|\frac{1}{N}\sum_{k}F^{(1)}(x_{m(k,t)}(0)+\theta_{k})\eta_{k}|\leq C(F,1)8C_{\alpha,g}\Delta
\]
Finally the bound is 
\[
C(F,2)\Delta^{2}+C(F,1)8C_{\alpha,g}\Delta
\]

To prove (\ref{lemma_fi_2}) is more difficult: one should obtain
maximal cancellation by carefully grouping the summation and integration
terms. Denote 
\[
b_{m}=\min(\Delta_{m-1},\Delta_{m}),I_{m}=(x_{m}(0)-\frac{b_{m}}{2},x_{m}(0)+\frac{b_{m}}{2})\subset S_{1}
\]
For any pair $(k,m)$ and any $t$ define the set $T(k,m)=T(k,m,t)\subset[0,t]$
as follows: $t_{1}\in T(k,m)$ iff 
\[
x_{k}(0)+Vt_{1}\in I_{m},t_{1}\leq t
\]
Note that for given $k$ the sets $T(k,m),m-1,...,N,$ are disjoint,
similarly for given $m$ the sets $T(k,m),k-1,...,N,$ are disjoint.
The set $T(k,m)$ consists of disjoint intervals 
\[
J_{i}=J_{i}(k,m),i=1,2,...,\beta=\beta(k,m,t),
\]
enumerated in the order of their hitting (imagine a particle starting
at $x_{k}(0)$ and moving with constant speed $V$), and $\beta=[tV${]}+1
or $\beta=[tV]+2$. All these intervals have length $|J_{i}|=V^{-1}b_{m}$,
except $J_{1}(k,k)$, having length $\frac{1}{2}V^{-1}b_{k}$ (we
call them initial intervals), and, for given $k$, possibly one of
the others (namely $J_{\beta}$, we call them end intervals), which
length can be less than $V^{-1}b_{m}$. We call the intervals $J_{i}(k,m)$,
having length $\frac{b_{m}}{V}$, regular, the others - non-regular.

\begin{lemma}\label{lemma_zeroMode_measure}

The sets $T_{1}(k)=[0,t]\setminus\cup_{m}T(k,m),k=1,...,N,$ and $T_{2}(m)=[0,t]\setminus\cup_{k}T(k,m),m=1,...,N,$
have measure less than $2(t+1)C_{\alpha,g}\Delta^{2}$.

\end{lemma}

In fact, from (\ref{delta_k}) it follows that 
\[
\delta_{m}-\delta_{m-1}=\sum_{l=1}^{\infty}a_{l}(Q_{m}^{l}-Q_{m-1}^{l})=-\alpha^{-1}\Delta^{2}(gF(x_{m})-w)+\sum_{l=2}^{\infty}a_{l}(Q_{m}^{l}-Q_{m-1}^{l})
\]

\begin{equation}
|\delta_{m}-\delta_{m-1}|\leq2C_{\alpha,g}\Delta^{2}\label{diff_delta_k}
\end{equation}
Then the assertion follows from 
\[
|\Delta_{m}-\Delta_{m-1}|=\Delta|\delta_{m}-\delta_{m-1}|\leq2C_{\alpha,g}\Delta^{3},\sum_{m=1}^{N}|\Delta_{m}-\Delta_{m-1}|\leq2C_{\alpha,g}\Delta^{2}
\]

Lemma \ref{lemma_zeroMode_measure} is proved.

We can write 
\[
\int_{0}^{t}\tilde{\phi}(0,t_{1})dt_{1}=\int_{0}^{t}dt_{1}\frac{1}{N}\sum_{k=1}^{N}(F(x_{k}(0)+Vt_{1})-F(x_{k}(0)))=\frac{1}{N}\int_{0}^{t}dt_{1}(\sum_{k=1}^{N}F(x_{k}(0)+Vt_{1})-\sum_{m=1}^{N}F(x_{m}(0)))=
\]

\begin{equation}
=\frac{1}{N}\sum_{k=1}^{N}(\sum_{m=1}^{N}\sum_{i=1}^{\beta}\int_{J_{i}(k,m)}+\int_{T_{1}(k)})F(x_{k}(0)+Vt_{1})dt_{1}-\frac{1}{N}\sum_{m=1}^{N}tF(x_{m}(0))\label{expansion_integral}
\end{equation}
and for any $m$ 
\[
tF(x_{m}(0))=F(x_{m}(0))(\sum_{k=1}^{N}\sum_{i=1}^{\beta}|J_{i}(k,m)|+|T_{2}(m)|)
\]
Take one of the regular intervals $J_{0}=J_{i}(k,m)$, then there
exists $t_{0}\in J_{0}$ such that $x_{k}(0)+Vt_{0}=x_{m}(0)$. Expanding
in $V(t_{1}-t_{0})$ we have

\begin{equation}
F(x_{k}(0)+Vt_{1})=F(x_{k}(0)+Vt_{0})+F'(x_{k}(0)+Vt_{0})V(t_{1}-t_{0})+F''(x_{k}(0)+Vt_{0}+\vartheta)V^{2}(t_{1}-t_{0})^{2}\label{expansion_J_0}
\end{equation}
for some $0\leq\vartheta\leq V(t_{1}-t_{0})$. After integration over
$J_{0}$ the linear term in the right-hand side of (\ref{expansion_J_0})
vanishes due to symmetry w. r. t. $t_{0}$, constant terms cancel
with the corresponding terms in the last sum in (\ref{expansion_integral}),
and as a result we get 
\[
|\int_{J_{0}}(F(x_{k}(0)+Vt_{1})-F(x_{m}(0)))dt_{1}|\leq C(F,2)\Delta^{3}
\]
The sum of such terms (for given $k$) over $m$ and over all $J_{i}(k,m)$
has the upper bound 
\begin{equation}
C(F,2)\Delta^{2}t\label{bound_regular}
\end{equation}
Taking into account the factor $\frac{1}{N}$, we have the same bound
after summation over $k$.

Consider now non-regular intervals, For the initial intervals we write
the expansion similar to (\ref{expansion_J_0}), taking $t_{0}=0$.
Then as above, the constant terms cancel and the linear terms, after
integration give bound $C(F,1)\Delta^{2}$. The end intervals $J_{\beta}$,
if there is some $x_{m}(0)\in J_{\beta}$, are treated similar and
give the same bound. If there are no such $x_{m}(0)\in J_{\beta}$
then we consider the union $J_{\beta-1}\cup J_{\beta}$ and do the
same procedure. As there are not more than $2N$ such intervals, the
bound will not depend on $t$. Namely. this give the bound 
\begin{equation}
\frac{1}{N}2NC(F,1)\Delta^{2}\label{bound_nonregular}
\end{equation}
Taking all together, namely Lemma \ref{lemma_zeroMode_measure}, (\ref{bound_regular})
and (\ref{bound_nonregular}), we get (\ref{lemma_fi_2}).

\section{Nonlinear integral equations}

We shall prove the Theorem by taking into account linear and non-linear
terms which we skipped in the basic linear approximation (\ref{zeroLinear}).
Remind that we assume for simplicity $u_{k}(0)=0$ and thus $C_{l}(n)-0$.
We consider $L_{j}=L_{j}(k,t)$ as functions of $k$ and $t$, given
$y_{k}(t)$ and denote 
\[
H(n,t)==\Phi(y_{k}(t)),\tilde{L}_{j}=\tilde{L}_{j}(n,t)=\Phi(L_{j}(k,t))
\]
(we will write down them explicitely below). Applying Fourier transform
to (\ref{main_nonlinear}) we get the main system of equations in
Fourier form 
\[
\frac{d^{2}H(n,t)}{dt^{2}}+\frac{A}{M}\frac{dH(n,t)}{dt}+\frac{4\alpha}{M}\Delta^{-3}(1-\cos2\pi\frac{n}{N})H(n,t)=M^{-1}\tilde{\phi}(n,t)+M^{-1}\sum_{j=1}^{3}\tilde{L}_{j}(n,t)
\]
Similarly to (\ref{eta_0}) we get the system of integral equations
for $H(n,t)$

\begin{equation}
H=K(H)+\eta\label{main_nonlinear_Fourier}
\end{equation}
with the non-linear integral operators, acting on $H$, 
\[
K=\sum_{j=1}^{3}K_{j},K_{j}=\sum_{l=1,2}K_{j,l},K_{j,l}(H)(n,t)=r_{l}(n)\int_{0}^{t}e^{z_{l}(n)(t-t_{1})}\tilde{L}_{j}(n,t_{1})dt_{1}
\]
and the free term (that we have studied above) 
\[
\eta(n,t)=\sum_{l=1,2}r_{l}(n)\int_{0}^{t}e^{z_{l}(n)(t-t_{1})}g\tilde{\phi}(n,t_{1})dt_{1}
\]

Define the Banach space $\mathbf{B}_{T}=\mathbf{B}_{T,N}$ of complex
(continuous in $t$) functions $b=b(n,t)$ on $\{0,1,...,N-1\}\times[0,T]$
with the norm 
\[
||b||=\sup_{n,t\in[0,T]}D_{n}|b(n,t)|
\]
where 
\[
D_{0}=\frac{1}{\Delta},
\]
and for $n\neq0$

\[
D_{n}=n\sqrt{\frac{\alpha}{M\Delta}}
\]
It follows that for any $n$ 
\begin{equation}
\sup_{t\in[0,T]}|b(n,t)|\leq||b||D_{n}^{-1}\label{sup_b_n_t}
\end{equation}
and thus for any subset $Q\subset\{0,1,...,N-1\}$ 
\begin{equation}
\sum_{n\in Q}\sup_{t\in[0,T]}|b(n,t)|\leq||b||\sum_{n\in Q}D_{n}^{-1}\label{sum_sup_b_n_t}
\end{equation}
Note that 
\begin{equation}
\sum_{n=1}^{N-1}D_{n}^{-1}\leq c\sqrt{\frac{M\Delta}{\alpha}}\ln N\label{sum_D_minus_1}
\end{equation}
Further on we put for any $0<\beta<1$ not depending on $N$ 
\[
T=T(\beta)=\beta NC_{A,g}^{-1}\geq c\beta\rho^{-1}N^{3}
\]
We shall prove that Banach fixed point theorem defines the unique
solution in $\mathbf{B}_{T}$.

\begin{lemma}\label{lemma_contract}

There exists $\gamma>0$, not depending on $N$ and such that for
any sufficiently large $N$ the ball 
\[
D(T,\gamma)=\{H\in\mathbf{B}_{T}:||H||\leq\gamma\}
\]
is invariant with respect to any operator K,$K_{j},K_{j,l}$ and for
any $H_{1},H_{2}\in\mathbf{B}_{T}$ for some $q<1$

\begin{equation}
||K(H_{1})-K(H_{2})||\leq q||H_{1}-H_{2}||\label{contraction_general}
\end{equation}
\end{lemma}

In particular, it follows from Lemma \ref{lemma_eta-1}, that $\eta\in D(T,\gamma)$.
It follows that there exists unique solution and we can solve the
equation (\ref{main_nonlinear_Fourier}) by the standard iteration.
We will prove Lemma separately for each $K_{j,l}$ with sufficiently
small $q$'s.

\paragraph{Bounds for $L_{1}$}

Consider two function $y_{k}(t)$ and $z_{k}(t)$ of $k$ and their
Fourier transforms 
\[
H_{1}(n,t)=\Phi(y_{k}(t)),H_{2}(n,t)=\Phi(z_{k}(t))
\]
Denote 
\[
u_{k}(y_{k})=F(x_{k}(0)+Vt+y_{k})
\]
Using Taylor expansion we have for some $\theta_{k}\in[z_{k},y_{k}],\theta'_{k}\in[0,z_{k}]$
\[
u_{k}(y_{k})-u_{k}(z_{k})=u_{k}(z_{k}+(y_{k}-z_{k}))-u_{k}(z_{k})=u_{k}^{(1)}(z_{k})(y_{k}-z_{k})+\frac{1}{2}u_{k}^{(2)}(z_{k}+\theta_{k})(y_{k}-z_{k})^{2}=
\]

\begin{equation}
=u_{k}^{(1)}(0)(y_{k}-z_{k})+z_{k}u_{k}^{(2)}(\theta'_{k})(y_{k}-z_{k})+\frac{1}{2}u_{k}^{(2)}(z_{k}+\theta_{k})(y_{k}-z_{k})^{2}\label{Taylor_u}
\end{equation}

We shall consider firstly the linear term 
\[
u_{k}^{(1)}(0)(y_{k}-z_{k})=F^{(1)}(x_{k}(0)+Vt)(y_{k}-z_{k})
\]
Using the convolution formula 
\[
\Phi(g_{1}g_{2},n)=(\Phi(g_{1})\star\Phi(g_{2}))(n)=\sum_{n_{1}}\Phi(g_{1},n-n_{1})\Phi(g_{2},n_{1})=\sum_{n_{1}}\Phi(g_{2},n-n_{1})\Phi(g_{1},n_{1})
\]
we have for $H=H_{1}-H_{2}$

\[
K_{1,l}(H)=gr_{l}(n)\int_{0}^{t}e^{z_{l}(n)(t-t_{1})}\sum_{n_{1}}\chi(n-n_{1},t_{1})H(n_{1},t_{1})dt_{1}
\]
where 
\[
\chi(n,t)=\frac{1}{N}\sum_{k=1}^{N}F^{(1)}(x_{k}(0)+Vt)\exp(2\pi in\frac{k}{N})
\]

\begin{lemma}\label{lemma_hi_} 
\begin{equation}
|\chi(0,t)|\leq\Delta C(F,2)\label{bound_hi_0}
\end{equation}
and for any $0<n\leq\frac{N}{2}$ 
\begin{equation}
|\chi(n,t)|\leq cn^{-1}\label{bound_hi_nonzero}
\end{equation}
Symmetrically for $\frac{N}{2}\leq n<N$.

\end{lemma}

Proof. As for any $t$ 
\[
\int_{S_{1}}F^{(1)}(x+Vt)dx=0
\]
we have

\[
|\chi(0,t)|=|\chi(0,t)-\int_{S_{1}}F^{(1)}(x+Vt)dx|=|\frac{1}{N}\sum_{k=1}^{N}F^{(1)}(x_{k}(0)+Vt)-\int_{S_{1}}F^{(1)}(x+Vt)dx|\leq\frac{1}{N}max|F^{(2)}(x)|
\]

For $n\neq0$ we can use the following summation-by-parts formula
\[
\sum_{k=0}^{N-1}h_{1}(k)\nabla h_{2}(k)=h_{1}(N)h_{2}(N)-h_{1}(0)h_{2}(0)-\sum_{k=0}^{N-1}h_{2}(k+1)\nabla h_{1}(k)
\]
where $h_{1}(N)=h_{1}(0),h_{2}(N)=h_{2}(0)$. Put in our case 
\[
h_{1}(k)=F^{(1)}(x_{k}(0)+Vt),h_{2}(k)=\sum_{l=0}^{k-1}\exp(2\pi in\frac{l}{N})=\frac{1-\exp(2\pi in\frac{k}{N})}{1-\exp(2\pi in\frac{1}{N})}
\]
Then 
\[
\nabla h_{2}(k)=\exp(2\pi in\frac{k}{N})
\]
If $n<\sigma N$, where for example $\sigma=\frac{1}{4\pi},$ then
\[
\frac{1}{N}|\sum_{k=0}^{N-1}h_{2}(k+1)\nabla h_{1}(k)|\leq\max_{k}|h_{2}(k+1)\nabla h_{1}(k)|\leq c\frac{N}{n}\frac{C(F,1)}{N}
\]
If $\sigma N\leq n\leq\frac{N}{2}$ then similarly 
\[
\frac{1}{N}|\sum_{k=0}^{N-1}h_{2}(k+1)\nabla h_{1}(k)|\leq\max_{k}|h_{2}(k+1)\nabla h_{1}(k)|\leq\sup_{\sigma\leq x\leq\frac{1}{2}}(|\frac{2}{1-\exp(2\pi ix)}|)\frac{C(F,1)}{N}=cC(F,1)n^{-1}
\]

To end the proof we shall prove a general assertion which will be
used also in further estimates.

\begin{lemma}\label{lemma_hi_general}

If the function $\chi(n,t)$ satisfies the bounds (\ref{bound_hi_0})
and (\ref{bound_hi_nonzero}) then under conditions (\ref{cond_main_1},\ref{cond_main_2})
the operator 
\[
K(H)=\sum_{l=1,2}gr_{l}(n)\int_{0}^{t}e^{z_{l}(n)(t-t_{1})}\sum_{n_{1}}\chi(n-n_{1},t_{1})H(n_{1},t_{1})dt_{1}
\]
satisfies the bound (\ref{contraction_general}) with $q$ sufficiently
small.

\end{lemma}

Proof. As in (\ref{eta_0}) we have by definition 
\[
||K_{1,l}(H)||\leq\sup_{n,t}D_{n}|gr_{l}(n)\int_{0}^{t}e^{z_{l}(n)(t-t_{1})}\sum_{n_{1}}H(n-n_{1},t_{1})\chi(n_{1},t_{1})dt_{1}|
\]
Firstly, for $l=1$ consider, under $\sup_{n}$, the case $n=0$,
namely 
\[
\sup_{t}D_{0}|gr_{l}(0)\int_{0}^{t}e^{z_{l}(0)(t-t_{1})}\sum_{n_{1}}H(-n_{1},t_{1})\chi(n_{1},t_{1})dt_{1}|\leq
\]

\[
\leq\Delta^{-1}g\frac{2}{A}\frac{M}{A}(\sum_{n_{1}\neq0}D_{n_{1}}^{-1}||H||\sup_{t_{1}}|\chi(n_{1},t_{1})|+D_{0}^{-1}||H||\sup_{t_{1}}|\chi(0,t_{1})|)\leq
\]

\[
\leq\Delta^{-1}2C_{A,g}\frac{M}{A}||H||(c\sqrt{\frac{M\Delta}{\alpha}}+C(F,2)\Delta^{2})\leq2C_{A,g}\frac{M}{A}||H||(c\sqrt{\frac{M}{\alpha\Delta}}\ln N+C(F,2)\Delta)
\]
For the case $n\neq0$ we have from (\ref{r_l}), similar to (\ref{eta_1_n_linear}),
\[
n\sqrt{\frac{\alpha}{M\Delta}}\frac{1}{n}\sqrt{\frac{M\Delta}{\alpha}}g\sum_{n_{1}\neq0}D_{n-n_{1}}^{-1}||H||\sup_{t_{1}}|\chi(n_{1},t_{1})|=g||H||\sum_{n_{1}\neq0}D_{n-n_{1}}^{-1}\sup_{t_{1}}|\chi(n_{1},t_{1})|
\]
The sum of the terms with $n_{1}\neq n$ does not exceed $c\sqrt{\frac{M\Delta}{\alpha}}\ln N$,
and the term with $n_{1}=n\neq0$ does not exceed $\frac{c}{n}\Delta.$
Finally we get the bound 
\[
c(\sqrt{\frac{M\Delta}{\alpha}}\ln N+\frac{1}{n}\Delta)g||H||
\]
If $l=2$, the case $n\neq0$ is similar to above. For the worst possible
term with $n=0,z_{2}(0)=0$ we have 
\[
\sup_{t}D_{0}|gr_{2}(0)||\int_{0}^{t}dt_{1}\sum_{n_{1}}H(-n_{1},t_{1})\chi(n_{1},t_{1})|\leq\Delta^{-1}\frac{2g}{A}t\sum_{n_{1}}\sup_{t}|\chi(-n_{1},t)H(n_{1},t)|\leq
\]

\[
\leq\Delta^{-1}\frac{2g}{A}t\sum_{n_{1}}\sup_{t}|\chi(-n_{1},t)|D_{n_{1}}^{-1}||H||
\]
The term 
\[
|\Delta^{-1}\frac{2g}{A}t\sup_{t}|\chi(-0,t)|D_{0}^{-1}||H||\leq2C_{A,g}C(F,2)t\Delta||H||
\]
with $n_{1}=0$ is small as 
\[
C_{A,g}\Delta t<c\beta
\]
The sum of the remaining terms (with $n_{1}\neq0$) 
\[
|\Delta^{-1}\frac{2g}{A}t\sum_{n_{1}\neq0}\sup_{t}|\chi(-n_{1},t)|D_{n_{1}}^{-1}||H||\leq2C_{A,g}\sum_{n_{1}}\frac{1}{n_{1}^{2}}\sqrt{\frac{M}{\alpha\Delta}}||H||
\]
is small by the right inequality of (\ref{cond_main_1}).

\pagebreak{}

The second (nonlinear) term 
\[
z_{k}u_{k}^{(2)}(\theta'_{k})(y_{k}-z_{k})
\]
is intuitively simpler because of the additional small factor $z_{k}$.
Using the Fourier transform $\chi_{1}\star H(n)$ is treated similarly
to the first term, where instead of $\chi(n,t)$ one should take 
\[
\chi_{1}(n,t)=\frac{1}{N}\sum_{k=1}^{N}z_{k}u_{k}^{(2)}(\theta'_{k})\exp(2\pi in\frac{k}{N})=\chi_{2}\star H_{2}
\]
where 
\[
\chi_{2}(n,t)=\frac{1}{N}\sum_{k=1}^{N}u_{k}^{(2)}(\theta'_{k})\exp(2\pi in\frac{k}{N})=\frac{1}{N}\sum_{k=1}^{N}F^{(2)}(x_{k}(0)+Vt+\theta'_{k})\exp(2\pi in\frac{k}{N})
\]
We will need only the obvious bound 
\begin{equation}
|\chi_{2}(n,t)|\leq C(F,2)\label{obvious_hi_2}
\end{equation}
but we can prove more. In fact, we have by (\ref{sum_sup_b_n_t})
and (\ref{sum_D_minus_1}) 
\begin{equation}
\sup_{t}|y_{k}(t)|\leq\sum_{n}\sup_{t\in[0,T]}|H_{1}(n,t)|\leq||H_{1}||\sum_{n}D_{n}^{-1}\leq||H_{1}||(\Delta+c\sqrt{\frac{M\Delta}{\alpha}}\ln N)\leq c(1+\rho)\Delta\label{y_k_less_Delta}
\end{equation}
as (\ref{cond_main_1}) gives 
\begin{equation}
\sqrt{\frac{M\Delta}{\alpha}}\ln N\leq c\rho\Delta\label{upper_bound_theorem}
\end{equation}
The same holds for $|z_{k}|$ and then also 
\[
|\theta_{k}|,|\theta'_{k}|\leq c\Delta
\]
This shows that the estimates of Lemma \ref{lemma_hi_} hold also
for $\chi_{2}(n,t)$, with the same proof.

Then by (\ref{sum_sup_b_n_t}) and (\ref{upper_bound_theorem}) 
\[
|\chi_{1}(n,t)|=|\chi_{2}\star H_{2}|(n,t)\leq C(F,2)\sum_{n}|H_{2}(n,t)|\leq C(F,2)||H_{2}||\sum D_{n}^{-1}
\]
Then using lemma \ref{lemma_hi_general}, we get the result. The third
term in (\ref{Taylor_u}) is treated similarly.

\paragraph{Bounds for $L_{2}$}

We have 
\[
L_{2}=2\alpha\Delta^{-3}\nabla^{-}[\delta_{k}^{1}\nabla y_{k}]=2\alpha\Delta^{-3}\delta_{k}^{1}\nabla^{-}\nabla y_{k}+2\alpha\Delta^{-3}(\nabla^{-}\delta_{k}^{1})(S^{-1}\nabla y_{k})
\]

\[
\tilde{L}_{2}(n,t)=2\alpha\Delta^{-3}\zeta_{1,2}\star\Phi(\nabla^{-}\nabla y_{k})+2\alpha\Delta^{-3}\zeta_{1,1}\star\Phi(S^{-1}\nabla y_{k}),
\]
where 
\[
\zeta_{1,2}(n)=\Phi(\delta^{1})=\frac{1}{N}\sum_{k=1}^{N}\delta_{k}^{1}\exp(2\pi in\frac{k}{N}),\zeta_{1,1}(n)=\Phi(\nabla^{-}\delta^{1})
\]
\begin{lemma}\label{lemma_hi_1_1}

For any $n$ 
\begin{equation}
|\zeta_{1,1}(n)|\leq c\alpha^{-1}g\Delta^{2}\label{bound_hi_1_1}
\end{equation}

\begin{equation}
|\zeta_{1,2}(n)|\leq c\alpha^{-1}g\Delta\label{bound_hi_1_2}
\end{equation}
\end{lemma}

This easily follows from Lemma \ref{lemma_w}, series (\ref{delta_k})
and bound (\ref{diff_delta_k}).

\begin{lemma}\label{lemma_hi_2}

For $q=1,2$ the operators 
\[
K_{2,l,q}(H)=r_{l}(n)\int_{0}^{t}e^{z_{l}(n)(t-t_{1})}\sum_{n_{1}}(\exp(-2\pi i\frac{n-n_{1}}{N})-1)^{q}H(n-n_{1},t_{1})\zeta_{1,q}(n_{1},t_{1})dt_{1}
\]
satisfy the bounds 
\[
||K_{2,l,q}(H)||\leq\alpha^{-1}g\Delta||H||
\]
\end{lemma}

Proof. We shall demonstrate the (straightforward) calculation for
$q=2$. The case $q=1$ is quite similar and even easier (because
of $\Delta^{2}$ factor).

For $l=1$, as in (\ref{eta_0}) we have by definition and by (\ref{derivatives_Fourier})
\[
||K_{2,1,2}(H)||\leq\sup_{n,t}D_{n}|r_{1}(n)\int_{0}^{t}e^{z_{1}(n)(t-t_{1})}\sum_{n_{1}}(\exp(-2\pi i\frac{n-n_{1}}{N})-1)^{2}H(n-n_{1},t_{1})\zeta_{1,2}(n_{1},t_{1})dt_{1}|
\]
Consider, under $\sup_{n}$, first the case $n=0$, namely 
\[
\sup_{t}D_{0}|r_{1}(0)\int_{0}^{t}e^{z_{1}(0)(t-t_{1})}\sum_{n_{1}}(\exp(2\pi i\frac{n_{1}}{N})-1)^{2}H(-n_{1},t_{1})\zeta_{1,2}(n_{1},t_{1})dt_{1}|\leq
\]

\[
\leq\Delta^{-1}\frac{2}{A}\frac{M}{A}(\sum_{n_{1}\neq0}D_{n_{1}}^{-1}||H||\sup_{t_{1}}\zeta_{1,2}(n_{1},t_{1})\leq c\frac{1}{A}\frac{M}{A}||H||\alpha^{-1}g\sqrt{\frac{M\Delta}{\alpha}}\ln N
\]

After multiplying on $\alpha\Delta^{-3}$ we get 
\[
c\alpha\Delta^{-3}\frac{1}{A}\frac{M}{A}||H||\alpha^{-1}g\sqrt{\frac{M\Delta}{\alpha}}\ln N\leq c\Delta^{-2}C_{A,g}\frac{M}{A}||H||\sqrt{\frac{M}{\alpha\Delta}}\ln N\leq c\rho||H||
\]
For the case $n\neq0$ we have, similar to (\ref{eta_1_n_linear}),
\[
\alpha\Delta^{-3}n\sqrt{\frac{\alpha}{M\Delta}}\frac{1}{n}\sqrt{\frac{M\Delta}{\alpha}}\sum_{n_{1}\neq0}D_{n-n_{1}}^{-1}||H||\sup_{t_{1}}|\zeta_{1,2}(n_{1},t_{1})|\leq
\]
\[
\leq\alpha\Delta^{-3}||H||\Delta cC_{\alpha,g}\Delta\leq cg\Delta^{-1}||H||
\]
Similarly for $l=2$ 
\[
||K_{2,2,q}(H)||\leq\sup_{n,t}D_{n}|r_{2}(n)\int_{0}^{t}e^{z_{2}(n)(t-t_{1})}\sum_{n_{1}}(\exp(-2\pi i\frac{n-n_{1}}{N})-1)^{2}H(n-n_{1},t_{1})\zeta_{1,2}(n_{1},t_{1})dt_{1}|
\]
The case $n\neq0$ is similar to above. For the worst possible term
with $n=0,z_{2}(0)=0$ we have,noting that the term with $n_{1}=0$
is zero, 
\[
\sup_{t}D_{0}|r_{2}(0)||t\sum_{n_{1}}(\exp(2\pi i\frac{n_{1}}{N})-1)^{2}H(-n_{1},t_{1})\zeta_{1,2}(n_{1},t_{1})|\leq\Delta^{-1}\frac{2}{A}t\sum_{n_{1}\neq0}\sup_{t}|\zeta_{1,2}(-n_{1},t)|D_{n_{1}}^{-1}||H||\leq
\]

\begin{equation}
\leq\Delta^{-1}\frac{2}{A}tc\alpha^{-1}g\Delta c\rho\Delta||H||\leq c\rho\frac{1}{A}t\alpha^{-1}g\Delta||H||\label{last_terrm}
\end{equation}
as 
\[
\sum_{n_{1}\neq0}D_{n_{1}}^{-1}\leq c\sqrt{\frac{M\Delta}{\alpha}}\ln N\leq c\Delta\sqrt{\frac{M}{\alpha\Delta}}\ln N\leq c\rho\Delta
\]
Finally 
\[
\alpha\Delta^{-3}c\rho\frac{1}{A}t\alpha^{-1}g\Delta||H||\leq c\rho t\Delta^{-2}C_{A,g}||H||\leq c\rho\beta||H||
\]
The last bound follows from (\ref{cond_main_2}). The lemma is proved.

\paragraph{Bounds for $L_{3}$}

We shall consider the $m$-th term $L_{3;m}$ of $L_{3}$ in the series
(\ref{L_3_series}) 
\[
\alpha(-\nabla^{-})\frac{(\nabla^{+}y_{k}(t))^{m}}{\Delta_{k}^{m+2}}=\alpha\frac{1}{\Delta_{k}^{m+2}}(-\nabla^{-})(\nabla^{+}y_{k}(t))^{m}+\alpha(S^{-1}\nabla^{+}y_{k}(t))^{m}(-\nabla^{-})\frac{1}{\Delta_{k}^{m+2}}
\]
The convergence of the series in $m$ will follow from the obtained
bound for $L_{3;m}$. Using the Leibnitz formula (\ref{Leibnitz})
we can rewrite the first term as follows 
\[
\alpha\frac{1}{\Delta_{k}^{m+2}}(-\nabla^{-})(\nabla^{+}y_{k}(t))^{m}=-\alpha\frac{1}{\Delta_{k}^{m+2}}\sum_{j=1}^{m}(\nabla^{+}y_{k}(t))^{j}(S^{-1}\nabla^{+}y_{k}(t))^{m-j-1}(\nabla^{-}\nabla^{+}y_{k}(t))
\]
Its Fourier transform will be 
\[
\Phi(\alpha\frac{1}{\Delta_{k}^{m+2}}(-\nabla^{-})(\nabla^{+}y_{k}(t))^{m})=\sum_{n_{1}}\zeta_{1}(n-n_{1},t)\Phi(\nabla^{-}\nabla^{+}y_{k}(t))(n)
\]
where 
\[
\zeta_{1}(n,t)=\Phi(-\alpha\frac{1}{\Delta_{k}^{m+2}}\sum_{j=1}^{m}(\nabla^{+}y_{k}(t))^{j}(S^{-1}\nabla^{+}y_{k}(t))^{m-j-1})
\]
Using Lemma \ref{lemma_y_k_0}, we have for $m\geq2$ 
\[
|\zeta_{1}(n,t)|\leq\alpha\Delta^{-m-2}m(cC_{A,g}\sqrt{\frac{M\Delta}{\alpha}}\ln N)^{m-1}\leq\alpha\Delta^{-m-2}m(c\rho\Delta^{4})^{m-1}\leq\alpha\Delta m(c\rho)^{m-1}
\]
Also by (\ref{y_eta_2}) 
\[
|\nabla^{-}\nabla^{+}y_{k}(t)|\leq\sum_{n\neq0}|H(n,t|\leq\sum_{n\neq0}D_{n}^{-1}||H||\leq c\rho\Delta||H||
\]
Then we have the result similarly to (\ref{last_terrm}). The second
term is treated similarly. It is interesting to note that nonlinear
terms demand less restrictive bound than (\ref{cond_main_2}).

\section{Comments}

There are many problems left. 
\begin{enumerate}
\item Most irritating and interesting is however only one: to include the
ionic lattice to the model of strongly interacting electrons. May
be a satisfactory picture can be obtained only on the quantum level.
However on the quantum level it is not clear even how to write down
the Schroedinger equation because the external field $F(x)$ is not
potential on the circle. 
\item With our methods we could not prove stability for any time $t\in(0,\infty)$
because of the zero mode problem, that is existence of zero root for
$n=0$. Additional linear term proportional to $y_{k}$ in the basic
equations (\ref{main_deviations}) could easily solve this problem
but I could not obtain this term as a result of realistic interaction
with the ionic lattice. 
\item Our assumption concerning smallness of $g$ is too restrictive at
least in two points. Firstly, if $\alpha^{-1}g$ is bounded then the
space scale $N^{-2}$ controls the effective forces acting on the
electrons. If $\alpha$ is smaller than $g$, then this scale will
be in-between $N^{-1}$ and $N^{-2}$, but when it becomes comparable
with the scale $N^{-1}$, then the macroscopic homogeneity will be
lost. In particular, the macro-velocity $V$ may depend on the distance
from the support of the external force. 
\item Secondly, the worst perturbation term is the linear term $L_{2}$.
Possibly, more refined techniques allow better estimates. \end{enumerate}

\end{document}